\documentclass[a4paper]{article}

\usepackage{INTERSPEECH2021,multirow,stfloats}
\usepackage{setspace}
\usepackage{threeparttable,booktabs,array}

\title{Phoneme-aware and Channel-wise Attentive Learning for Text Dependent Speaker Verification}
\name{Author Name$^1$, Co-author Name$^2$}
\name{Yan Liu $^1$, Zheng Li $^{1}$, Lin Li $^{1*}$, Qingyang Hong $^{2*}$ }
\address{$^1$School of Electronic Science and Engineering, Xiamen University, China \\ 
$^2$ School of Informatics, Xiamen University, China}
\email{lilin@xmu.edu.cn, qyhong@xmu.edu.cn}

\begin{document}

\maketitle
\begin{abstract}
This paper proposes a multi-task learning network with phoneme-aware and channel-wise attentive learning strategies for text-dependent Speaker Verification (SV). 
In the proposed structure, the frame-level multi-task learning along with the segment-level adversarial learning is adopted for speaker embedding extraction. 
The phoneme-aware attentive pooling is exploited on frame-level features in the main network for speaker classifier, with the corresponding posterior probability for the phoneme distribution in the auxiliary subnet. 
Further, the introduction of Squeeze and Excitation (SE-block) performs dynamic channel-wise feature recalibration, which improves the representational ability. 
The proposed method exploits speaker idiosyncrasies associated with pass-phrases, and is further improved by the phoneme-aware attentive pooling and SE-block from temporal and channel-wise aspects, respectively.
The experiments conducted on RSR2015 Part 1 database confirm that the proposed system achieves outstanding results for text-dependent SV.
\end{abstract}

\section{Introduction}
The aim of SV is to verify the client's claimed identity using his/her speech. Given the spoken contents used for enrollment and testing, SV can be categorized into text-dependent and text-independent \cite{628714}. The former task requires fixed content for the enrollment and testing speech, while the latter is more flexible, which does not need such constraints on speech content \cite{7746659}. 

Recently, neural networks have given SV a boost. Snyder et al. \cite{xv} proposed to aggregate variable-length utterances into fixed-dimension speaker embeddings, known as x-vector. As x-vector \cite{xv} has achieved great success in SV task, it has been optimized and replaced i-vector \cite{iv} as the baseline. In text-independent task, it generally requires a larger amount of training data, and the phonetic information should be wiped out in the final speaker embeddings, so text-independent techniques do not always suit for text-dependent scenarios where both speakers and texts are discriminated simultaneously. 

Text-dependent SV can be considered as a combination of two tasks, namely speaker verification and utterance verification, which can be optimized jointly or separately \cite{uttsv, spoken}. The importance of content modelling in text-dependent SV has been highlighted in many prior researches. Dey et al. \cite{dtw} used Dynamic Time Warping (DTW) to match the content and a Deep Neural Network-Hidden Markov Model (DNN-HMM) framework was utilized to extract phonetic posteriors. The DNN based multi-task approach extracted speaker-phoneme features to capture the lexical information for text-dependent SV \cite{dnn}. Aside from directly incorporating content information in training network, speaker-word pair was used to train the Probabilistic Linear Discriminant Analysis (PLDA) model so that the backend system took into account the text labels\cite{content,new3}.

Attention-based models have recently become popular as they have gained outstanding performance in SV tasks. In works like \cite{selfatt, ap}, an attention mechanism was applied in pooling in order to aggregate frame-level features selectively into speaker embeddings. Besides pooling, attention can also be used in network layers to address the problem of sequence summarization \cite{attsv}. In \cite{sesv}, the authors introduced the SE mechanism to explicitly model the interdependencies between the channel-wise speaker features. In \cite{new5}, the bidirectional attentive pooling models global temporal context and aggregates important frames which contains more speaker information.

Recently, more research has focused on schemas of phoneme-aware networks \cite{new2,new4}. In \cite{8070977}, a phonetic-discriminative DNN produced phonetic features which substituted raw acoustic features as the input to a Long Short-Term Memory Recurrent Neural Network (LSTM-RNN).
In \cite{hmmiv}, the monophone HMM models were concatenated into the phrase-specific HMMs that were used to collect sufficient statistics for i-vector. 
In \cite{9054333}, the text adaptation framework factorized utterances into speaker embedding and text embedding to address text mismatch issue.

In this paper, we wish to disentangle speaker and text information in text-dependent SV tasks, and we propose a multi-task framework with phoneme-aware attentive pooling and SE-block to improve the performance gain. The multi-task framework learns the speaker and phoneme information simultaneously, specifically, the frame-level multi-task learning along with the segment-level adversarial learning work as auxiliary subnets to improve the speaker discriminative capability with regard to different pass-phrases. Since the conventional speaker embedding uses statistics pooling to calculate the mean and standard deviation of all frame-level feature to form utterance-level feature, however, it cannot automatically select the most speaker-phoneme related frames, we propose a novel attentive pooling mechanism to utilize the phoneme posterior from the phonetic-discriminative subnet to generate weighted means and weighted standard deviations. From the temporal view, this attention mechanism  makes full use of phonetic information and benefits the long-term variations in speaker representation. 
This paper also integrates the SE-block to the multi-task framework, and to the best of our knowledge, it  has not been investigated for text-dependent SV.
Experimental results on Part \uppercase\expandafter{\romannumeral1} of RSR2015 corpus show the superiority of the proposed framework, where speaker embedding extraction benefits from shared information correlated with different tasks.

The rest of the paper is organized as follows. 
The related works are introduced in Section 2. 
Section 3 describes the proposed multi-task learning framework and phoneme-aware attentive pooling and SE-block for speaker verification. 
Section 4 details the experimental setup, followed by results and analysis in Section 5. The paper is finally concluded in Section 6.
\section{Related works}
\subsection{The frame-level multi-task learning}
\begin{figure}[h]
\centerline{\includegraphics[scale=0.8]{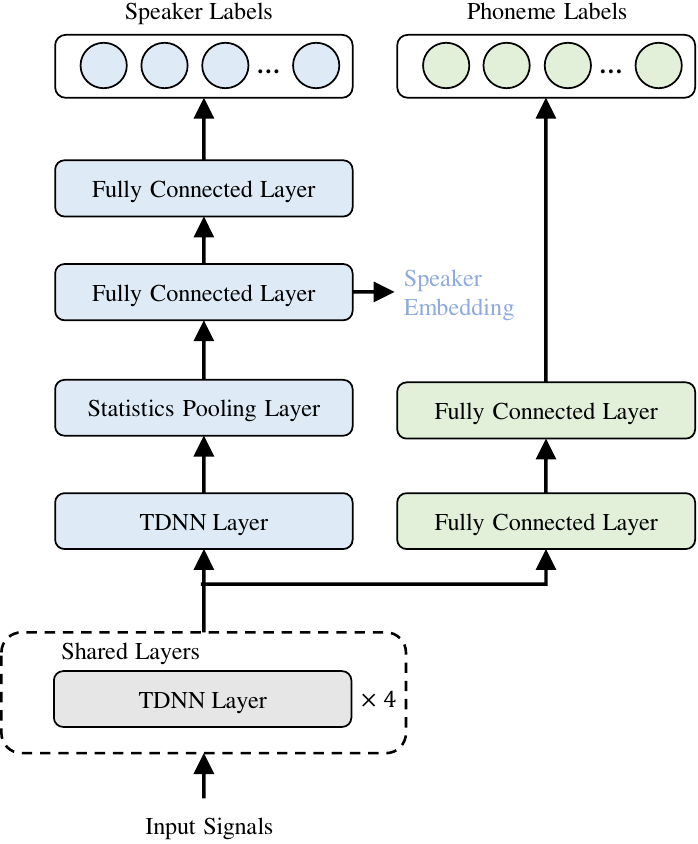}}
\caption{The frame-level multi-task learning.}
\label{figmf}
\end{figure}

Compared to single-task networks such as x-vector \cite{xv}, the frame-level multi-task architecture \cite{pho,dcl} introduces phonetic information to speaker embedding extraction as shown in Figure \ref{figmf}. The shared layers at the beginning of the network extract more informative features, which benefit the model generalization. The phonetic-discriminative subnet is trained at frame level while the speaker classification is done at segment level. 
\vspace{-0.5cm}
\subsection{The squeeze and excitation block}
The SE-block has been widely used in SV community. The output of the network layer $O \in R^{T*C}$ is the input of SE-block, where $C$ denotes the number of channels of the layer and $T$ refers to the frame number of the utterances. The statistics pooling layer first integrates features into channel-wise mean and standard deviation. Then the channel-wise dependencies are extracted by two fully-connected layers with nodes of $\frac{C}{r}$ and $C$ respectively, where $r$ denotes the reduction ratio which depends on the trade-off between the performance gain and computational cost \cite{se}. Finally, a scale operator is used for channel-wise multiplication to emphasize informative features and suppress less useful ones.


\section{The speaker-phoneme multi-task learning}
Extended Time-Delay Neural Network (E-TDNN) based architecture has been widely used as the speaker embedding extractor in SV tasks \cite{etdnn}. In this paper, we explore the phonetic multi-task learning structure based on E-TDNN. 

In conventional speaker embeddings, statistics pooling is usually used to integrate features from all the frames of a single utterance to form a segment-level representation with the same weight. However, phonetic information plays an important role in text-dependent SV. Thus, we incorporate phoneme-aware attentive pooling in the multi-task learning framework to improve discriminative learning capability of the embedding extraction. 

\begin{figure}[!h]

\centerline{\includegraphics[scale=0.8]{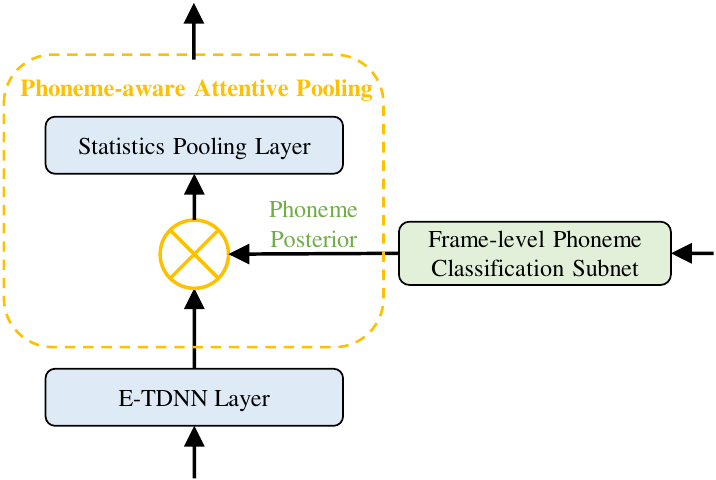}}

\caption{Details of the phoneme-aware attentive pooling.}
\label{figpho}
\end{figure}

The proposed pooling method is illustrated in Figure \ref{figpho}, and Equation (\ref{eq}) details the mechanism. $p$ is the phoneme posterior of the output of $M^{pf}$, and $out^{l5}$ denotes the output of the fifth E-TDNN layer, and $Pool$ indicates the statistics pooling. By multiplying the posterior from the frame-level phoneme-aware subnet to the representation from the speaker subnet, different weights are applied to the mean and standard deviation of the frame-level speaker representation, so that the speaker subnet will attain more direct and precise phonetic information. Inspired by \cite{2017attention}, the softmax function and scaling operator are applied to the dot-product in order to avoid the conditions where the dot products grow extremely small in magnitude, and $scale$ is set to 1.5.
\begin{equation}
PhoneAttPool=Pool(scale \cdot Softmax(p \cdot out^{l5}))
\label{eq}
\end{equation}
\vspace{-0.5cm}
\vspace{-0.3cm}

\begin{figure}[h]
\centerline{\includegraphics[width=7.9cm]{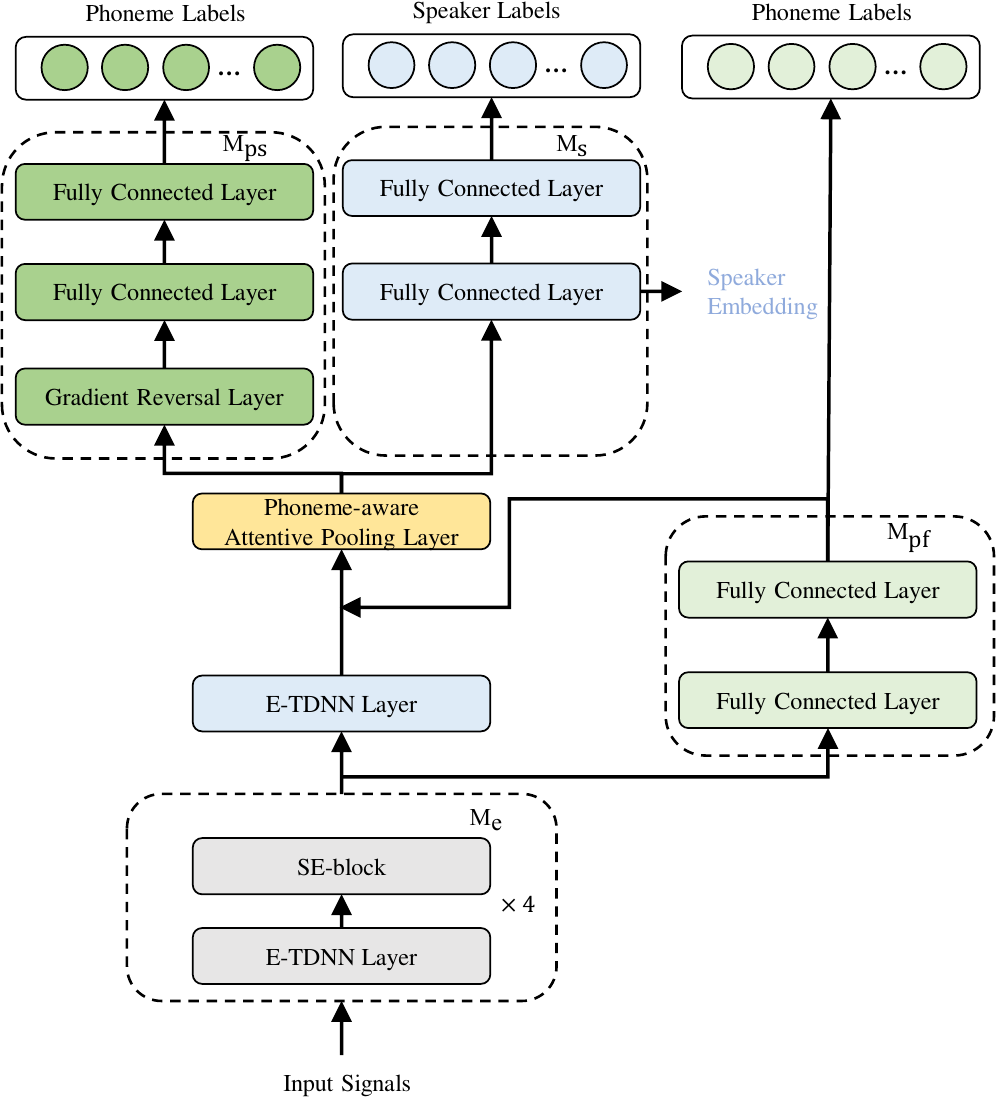}}
\caption{The multi-task learning network with phoneme-aware attentive pooling and SE-block.}
\label{figmt}
\end{figure}

The frame-level multi-task network is studied first and we substitute the TDNN layers with E-TDNN layers. However, this basic structure could not achieve a satisfying result, the segment-level adversarial structure \cite{dcl} is then combined with the original network to build a more robust architecture which is depicted in Figure \ref{figmt}. The network configuration is outlined in Table \ref{tabnet}.
\begin{table}[h]
\scriptsize
\caption{The proposed architecture. Speaker embeddings are extracted
at speaker subnet segment-level 1. The N and P in loss function
layers correspond to the number of training speakers and the number of phoneme labels, respectively. }
\label{tabnet}

\begin{tabular*}{7.9cm}{@{\extracolsep{\fill}}c|c|c}
\hline
Layer                           & Layer Context               & Input x Output \\ \hline
Shared Layer 1                  & {[}t-2, t-1, t, t+1, t+2{]} & 23 x 512       \\ \hline
Shared Layer 2                  & {[}t-2,t,t+2{]}             & 512 x 512      \\ \hline
Shared Layer 3                  & {[}t-3,t,t+3{]}             & 512 x 512      \\ \hline
Shared Layer 4                  & t                           & 512 x 512      \\ \hline
E-TDNN                          & t                           & 512 x P    \\ \hline
Phonetic Subnet Frame-level 1   & t                           & 512 x 512      \\ \hline
Phonetic Subnet Frame-level 2   & t                           & 512 x 512      \\ \hline
Phoneme-aware Attentive Pooling & {[}0,P)                     & P x 2P         \\ \hline
Speaker Subnet Segment-level 1  & 0                           & 2P x 512       \\ \hline
Speaker Subnet Segment-level 2  & 0                           & 512 x 512      \\ \hline
Phonetic Subnet GRL             & -                           & -              \\ \hline
Phonetic Subnet Segment-level 1 & 0                           & 2P x 512       \\ \hline
Phonetic Subnet Segment-level 2 & 0                           & 512 x 512      \\ \hline
Speaker Subnet Loss             & 0                           & 512 x N        \\ \hline
Frame-level Phonetic Subnet Loss & 0 & 512 x P \\ \hline
Segment-level Phonetic Subnet Loss & 0 & 512 x P \\ \hline
\end{tabular*}

\end{table}

The framework is composed of two main modules, including a frame-level phoneme-aware multi-task learning and a segment-level adversarial learning with phonetic information. The multi-task learning structure includes a shared frame-level feature extractor $M_e$, two parallel subnets $M_s$ and $M_{pf}$ for segment-level speaker classification and frame-level phoneme discrimination learning, respectively. For the feature extractor $M_e$, the basic architecture contains four E-TDNN layers. The SE-block is inserted behind each E-TDNN layer in $M_e$. As for the segment-level adversarial learning structure $M_{ps}$, the segment-level phonetic information can be suppressed by a Gradient Reversal Layer (GRL) \cite{grl}. In the GRL, the forward propagation in the network remains the same as other hidden layers, while the gradients are reversed in the backward propagation, so it achieves a kind of adversarial training. 

For a segment with $N$ frames, $X=\{x_1,x_2, …, x_N\}$, the corresponding speaker label is $y^s$, the frame-level phoneme label is $y^{pf}$, and the segment-level phoneme label is $y^{ps}$. The total loss of the speaker-phoneme multi-task network is defined as $L_{total} = L_s + \alpha L_{pf} + \beta L_{ps}$, where $L_s$ , $L_{pf}$, and $L_{ps}$ denote the loss functions of speaker classification subnet, frame-level phonetic multi-task subnet, and segment-level GRL subnet, respectively. $\alpha$ and $\beta$ are the weights of $L_{pf}$ and $L_{ps}$, and are set to 0.3 and 0.2, respectively. $CE$ denotes the cross entropy loss, and $KL$ denotes the Kullback-Leibler loss. Since the phoneme alignment corresponds to each frame rather than each segment and the phonetic labels at segment level are soft labels in the form of probabilities rather than the indices of specific phonemes, $KL$ loss substitutes $CE$ loss for the segment-level network.
\begin{equation}
L_s = CE(M_s(M_e(X)), y^s)
\end{equation}
\begin{equation}
L_{pf} = \frac{1}{N} \sum_{i=1}^{N} CE(M_{pf}(M_e(x_i)), y_i^{pf})
\end{equation}
\begin{equation}
L_{ps} = \frac{1}{N} \sum_{i=1}^{N} KL(M_{ps}(M_e(x_i)), y_i^{ps})
\end{equation}

\section{Experiments}
\label{sec:pagestyle}
\subsection{Datasets}
The Part \uppercase\expandafter{\romannumeral1} of RSR2015 corpus contains 16kHz English speech from 300 speakers (143 females, 157 males), which includes 30 fixed phrase utterances of 3-4 seconds duration. Each phrase is spoken 9 times by all the speakers. The first, fourth, and seventh sessions, recorded on Device A, are used for speaker enrollment, and the remaining recorded on Device B and C are used for testing \cite{rsr}. The Part \uppercase\expandafter{\romannumeral1} is divided into three subsets: background, development, and evaluation. The background subset is used for training, and the rest for enrollment and trial evaluation. 

For a standard text-dependent task, the test trials are grouped under four types based on the speaker and phrase labels as shown in Table \ref{table2015}, among which TC denotes the target trials, and the remaining three categories represent non-target trials. We report the performance of our models in terms of Equal Error Rate (EER).
 
\begin{table}[h]
\begin{center}
\caption{Types of trials for text-dependent SV}
\begin{spacing}{1.19}
\label{table2015}
\begin{tabular}{ccc p{8cm}}
\hline
 & Correct Content & Wrong Content\\ \hline
Target & TAR-correct (TC)& TAR-wrong (TW) \\
 Impostor & IMP-correct (IC)& IMP-wrong (IW) \\ 
 \hline
\end{tabular}
\end{spacing}
\end{center}
\vspace{-1cm}
\end{table}

\subsection{Experimental setup}
According to the Kaldi recipe \cite{kaldi}, a triphone GMM based Automatic Speaker Recognition (ASR) model is trained based on the background data, including 1708 Probability Density Function Identification (PDF-IDs). In order to get more accurate phoneme alignment, the PDF-IDs for each frame are used to represent the corresponding phoneme labels.

The i-vector, HiLAM \cite{rsr}, and the standard TDNN based x-vector \cite{xv} are taken as three baselines in this paper. All the proposed architectures are modified from the x-vector system. In the testing phase, only the speaker classification subnet is remained in which the penultimate layer at segment level is extracted as the speaker embeddings. 
\begin{center}
\begin{table}[h]

\vspace{-0.4cm}
\center
\setlength{\tabcolsep}{1.0mm}
\begin{spacing}{1.19}
\caption{Performance in EER (\%) for pre-experiments on RSR2015 Part \uppercase\expandafter{\romannumeral1} development set.}
\label{table2}
\begin{threeparttable}
\setlength{\tabcolsep}{1.5mm}{
\begin{tabular}{ccccccc}
\hline
\multirow{3}{*}{System}  & \multicolumn{6}{c}{Development set} \\
\cline{2-7}
 & \multicolumn{3}{c}{Male}& \multicolumn{3}{c}{Female}  \\
\cline{2-7}
 & TW & IC & IW & TW & IC & IW  \\ 
 \hline
 S1\tnote{1} & 1.534 & 1.545 & 0.112 & 0.677 & 1.223 & 0.036  \\
 S2\tnote{2}& 1.053  & 1.545 & 0.100 & 0.605 & 1.271  & 0.030  \\
 S3\tnote{3}&  1.467 & 1.601 & 0.089 & 0.502 & 1.223 &   0.028 \\
 S4\tnote{4}& \textbf{ 0.694 } & 1.650 & \textbf{0.056}  &\textbf{ 0.403} & 1.256 & \textbf{0.024}  \\
\hline
\end{tabular}
}
\begin{tablenotes}
\footnotesize
\item [1] Frame-level multi-task with statistics pooling.
\item [2] Frame-level multi-task with statistics pooling and SE-block.
\item [3] Frame-level multi-task with phoneme-aware attentive pooling.
\item [4] Frame-level multi-task with phoneme-aware attentive pooling and SE-block.
\end{tablenotes}
\end{threeparttable}
\end{spacing}
\end{table}
\end{center}
\vspace{-0.4cm}
\par The speech utterances are processed with 25ms frame length and 10ms shift to extract 20-dimensional Mel-frequency Cepstral Coefficient (MFCC) with 3-dimensional pitch feature. Cepstral Mean Normalization (CMN) with a sliding window of up to 3 seconds and Voice Activity Detection (VAD) based on short-time energy and zero crossing rate are applied to the extracted features. No data augmentation is applied in all experiments.

The proposed models for speaker embedding extraction are implemented on the open-source speaker and language recognition toolkit ASV-Subtools \cite{asvtools}, which is developed based on Pytorch \cite{py}. In order to investigate the impacts of phoneme-aware attentive pooling and SE-block on the multi-task networks, we also conduct a set of experiments without these two modules separately. The models are optimized with Adam optimizer, with a mini-batch size of 256 and chunk size of 100 in each segment. For the back-end, the background set is used for training the PLDA models with strategies including centering, whitening, and normalization. The acoustic feature extraction and back-end process are executed on Kaldi \cite{kaldi}.


\section{Results and analysis}
As the work aims at multi-task learning, joint speaker utterance (joint-spk-utt) \cite{joint} and unified speaker and utterance verification (Unified SUV) \cite{joint2} are considered for comparison. It is noticed that the joint-spk-utt models the speaker and speech information jointly \cite{joint}, whereas the Unified SUV uses a shared LSTM to drive two separate stacks of LSTM layers for speaker verification and utterance verification, respectively. The E-TDNN based standard x-vector \cite{xv} with statistics pooling and SE-block (S5) is also used as a reference system. 
\vspace{-0.5cm}
\begin{center}
\begin{table*}[ht]
\center
\setlength{\tabcolsep}{2.1mm}
\begin{spacing}{1.19}
\caption{Performance in EER (\%) for proposed system in comparison to existing systems on RSR2015 Part \uppercase\expandafter{\romannumeral1} corpus.}
\label{table1}
\begin{threeparttable}
\begin{tabular}{ccccccccccccc}
\hline
\multirow{3}{*}{System}  & \multicolumn{6}{c}{Male} & \multicolumn{6}{c}{Female}\\
\cline{2-13}
 & \multicolumn{3}{c}{Development}& \multicolumn{3}{c}{Evaluation}  & \multicolumn{3}{c}{Development}& \multicolumn{3}{c}{Evaluation} \\
\cline{2-13}
 & TW & IC & IW & TW & IC & IW  & TW & IC & IW  & TW & IC & IW \\ 
 \hline
i-vector \cite{rsr} & 2.870 & 5.950 & 0.740 & 1.950 & 4.030 & 0.320 & 3.050 & 7.870 & 0.940 & 1.910 & 6.610 & 0.750 \\
HiLAM \cite{rsr} & 1.660 & 3.690 & 0.490 & 0.820 & 2.470 & 0.190 & 1.770 & 3.240 & 0.450 & 0.610 & 2.960 & 0.140 \\

Joint-spk-utt \cite{joint} & 5.565 & 1.981 & 1.792 & 5.125 & 2.079 & 0.888 & 5.179 & 1.699 & 0.831 & 3.110 & 1.453 & 0.499 \\
Unified SUV \cite{joint2} & 0.470 & 1.590 & 0.101 & 0.293 & 1.757 & 0.039 & 1.176 &4.323 &0.178 &0.375 &2.009 &0.068 \\
\hline
 S4 \tnote{1} & 0.694 & 1.650 &  0.056  & 0.498 & 1.162 & 0.029 & 0.403 & 1.256 & 0.024 & 0.204 & 1.044 & 0.023 \\
 S5 \tnote{2} &5.509  & 2.318   & 0.492 & 3.017 & 2.067  & 0.285 & 5.310 & 2.499 & 0.566 &  2.337 & 2.02 & 0.283 \\

 S6 \tnote{3}  & 0.492 & \textbf{1.579} & \textbf{0.033}  & \textbf{0.237} & 1.212 & \textbf{0.023} & \textbf{0.312} & \textbf{1.103} & \textbf{0.019} & \textbf{0.124} & \textbf{0.874} & \textbf{0.022} \\


\hline
\end{tabular}
\begin{tablenotes}
\footnotesize
\item [1] Frame-level multi-task with phoneme-aware attentive pooling and SE-block.
\item [2] E-TDNN based x-vector with statistics pooling and SE-block.
\item [3] Frame-level multi-task and segment-level adversarial learning with phoneme-aware attentive pooling and SE-block.
\end{tablenotes}
\end{threeparttable}
\end{spacing}
\end{table*}
\end{center}
\vspace{-3pt}
\par We first conduct a sets of experiments on the development set to investigate how the phoneme-aware attentive pooling and SE-block will impact the frame-level multi-task network. As shown in Table \ref{table2}, both the phoneme-aware attentive pooling and SE-block improve the frame-level multi-task network with statistics pooling (S1) on TW and IW trials. The phoneme-aware attentive pooling (S3) obtains better results than SE-block (S2) on IW trials. The multi-task network with phoneme-aware attentive pooling and SE-block (S4) outperforms the above three systems except for IC trials. The results show that the SE-block improves the system performance, especially for RSR2015 testing conditions where the enrollment and test sets are cross-channel recorded. The phoneme-aware attentive pooling directly utilizes the phonetic information learned at frame level and is proved to be effective for TW and IW trials.

Table \ref{table1} shows the comparison of our proposed system (S6) to the systems mentioned above on Part \uppercase\expandafter{\romannumeral1} of RSR2015 corpus. We notice that the GMM-HMM based HiLAM outperforms the i-vector system due to the fact that the former system models the temporal information. For the joint-spk-utt model, it performs better than HiLAM system in the IC trial, but it degenerates in the utterance-related TW and IW trials. 
The S5 significantly outperforms the baselines for IC and IW trials; however, the x-vector does not model the lexical information which is of critical importance in text-dependent SV tasks. 
Further, the Unified SUV which utilizes LSTM model to extract temporal information further enhances the performance, especially for TW and IW trials. 



The results from two multi-task systems demonstrate that the introduction of phonetic information into text-dependent SV models obviously improves the basic E-TDNN network. The combination of frame-level phoneme-aware multi-task learning and the segment-level adversarial training with phoneme-aware attentive pooling and SE-block (S6) achieves the best performance, which indicates that the multi-task system benefits from the phoneme-enhancement at frame-level and phoneme-suppression at segment level, and the introduction of phonetic information to the pooling module and total loss function significantly improves the text-dependent SV performance. Additionally, the SE-block recalibrates the channel-wise features and improves the systems by capturing subtle differences between identities.


\section{Conclusions}
In this paper, we have explored approaches to improve the text-dependent SV performance from three aspects: phonetic loss auxiliary, temporal attentive aggregation, and channel-wise feature recalibration. 
The multi-task learning network with phoneme-aware attentive pooling and SE-block learns speaker-phoneme features with informative features reinforced and less useful ones suppressed.
Results showed that the proposed system could achieve outstanding performance on Part \uppercase\expandafter{\romannumeral1} of RSR2015 corpus. The future work will focus on optimizing the pooling mechanism to further improve the performance.

\section{Acknowledgements}
This work is supported by the National Natural Science Foundation of China (Grant No. 61876160 and Grant No. 62001405).

\bibliographystyle{IEEEtran}
\bibliography{refs}


\end{document}